\documentstyle[12pt]{article}
\begin{document}
  
  \centerline{\normalsize\bf EXTREMELY HIGH ENERGY COSMIC RAYS AND}
  \baselineskip=16pt
  \centerline{\normalsize\bf THE AUGER OBSERVATORY}
  \vspace*{0.6cm}
  \centerline{\footnotesize MURAT BORATAV}
  \baselineskip=13pt
  \centerline{\footnotesize\it LPNHE, Universit\'e Paris 6, 4 Place Jussieu}
  \baselineskip=12pt
  \centerline{\footnotesize\it 75005 Paris, France}
  \centerline{\footnotesize E-mail: boratav@in2p3.fr}
  \vspace*{0.9cm}
  \begin{quote}
  {\footnotesize\centerline{ABSTRACT}
   Over the last 30 years or so, a handful of events observed in
  ground-based cosmic ray detectors seem to have opened a new window in the
  field of high-energy astrophysics. These events have energies exceeding
  $5\times10^{19}$ eV (the region of the so-called Greisen-Zatsepin-Kuzmin
  spectral cutoff)~; they seem to come from no known astrophysical source~; 
  their chemical composition is mostly unknown~; no conventional accelerating
  mechanism is considered as being able to explain their production and
  propagation to earth. Only a dedicated detector can bring in the
  high-quality and statistically significant data needed to solve this
  long-lasting puzzle~: this is the aim of the Auger Observatory project around
  which a world-wide collaboration is being mobilized.}
  \end{quote}
  \normalsize\baselineskip=15pt
  \setcounter{footnote}{0}

  \section{Introduction}
This article will mainly concern the problems raised by the existence and
observation of cosmic rays whose energies are around and above $10^{20}$ eV (or
100 EeV). Such cosmic rays -that we shall call the ``extremely high energy
cosmic rays" or EHECR- are exceptional for the following reasons~:

\begin{itemize}
\item Even when one takes into account the experimental uncertainties on their
energy measurement (of the order of 20\%), they are well above the so-called
Greisen-Zatsepin-Kuzmin (GZK) cutoff. This cutoff corresponds to the threshold
for inelastic collisions between the cosmic microwave background (CMB) and
protons (photo-pion production) or heavy nuclei (photo-disintegration). The
result is that no EHECR observed on earth can be produced by sources very
distant from us. We shall come later on this very important point.

\item There are very few (if none) conventional astrophysical sources considered
unanimously by the experts as being able to accelerate particles at energies
exceeding those of the most energetic of the EHECRs that have been observed in
the past. 

\item At such energies, the effect of the galactic and extra-galactic magnetic
fields are quite weak on the EHECRs. Thus, even if those are charged particles,
their reconstructed incident direction should point toward their sources within
a few degrees. This distinguishes the EHECRs from their brothers in the lower
energy regions~: one can use them for point-like-source-search astronomy.
\end{itemize}

In the following, we shall try to demonstrate that if one takes into account all
the necessary conditions for an EHECR to be observed on earth (production 
energy, production rate, escape from production site, propagation, interaction 
with earth's atmosphere), the natural conclusion should be that the EHECRs do 
not exist. However, they exist and have been detected, although in small 
numbers~: less than 50 events above 40 EeV, and only 9 with energies equal to 
or larger than
100 EeV. This is a very exciting situation that one could actually 
consider as
being symmetrical with the present status of the dark matter problem~: in one
case (the dark matter) we deal with ``particles which {\it should} exist and 
don't, while in the other (the EHECRs) with particles which {\it do} exist but
perhaps shouldn't" \cite{Celnikier}.

One last remark about the relevance of the Auger Observatory with the main topic
of this conference~: the neutrino telescopes. Other contributions \cite{Halzen}
showed during this meeting that the detection of neutrino induced showers in the
atmosphere could be a very interesting alternative to other techniques
(telescopes immerged in water or ice) at the highest energies. On top of that,
no decisive argument can rule out a neutrino component for the EHECRs~;
actually, the neutrino hypothesis would make life easier when looking for a
solution to the EHECR puzzle (no upper limit would be necessary for the
distance of the source) although this would be at the cost of raising a new
series of problems. We shall come back shortly on this aspect in the following.

\section{What do we know of the EHECRs?}
The first -and the most important- fact on these cosmic rays is that they exist!
If we concentrate on those with energies equal to, or larger than, $10^{20}$ eV
(this is an arbitrary threshold, no physical reason justifies this special
value), 9 events were observed in the past, with five different ground
detectors~: Volcano Ranch \cite{Linsley}, Haverah Park \cite{Lawrence}, 
Yakutsk \cite{Afanasiev}, Fly's Eye \cite{Bird} and AGASA \cite{Yoshida}. The
detections were spread over more than 30 years and the detectors involved are
all based on different detection techniques (scintillators with or without muon
detection, air- and water-Cerenkov, atmospheric fluorescence). Thus, it is
almost impossible to envisage that the detected events could result from
experimental biases, artefacts or calibration errors.

The flux of these cosmic rays is extremely low at the energies we are interested
in. This can be roughly parametrized by a simple law~:

$$I(E>E_0)=\frac{100}{E_0^2}$$
in km$^{-2}$.sterad$^{-1}$.year$^{-1}$ when $E_0$ is given in EeV. The two most
recent (and most energetic) of these events have energies of 200 and 320 EeV,
with uncertainties of less than 30\% (including systematics).

Nothing decisive can be said on the chemical composition on these events. The
only detector capable of providing a somewhat precise information on this
feature is the Fly's Eye. Their data including the lower energy region of the
spectrum is compatible with the interpretation that between 1 and 10 EeV the
chemical composition of the cosmic rays shifts gradually from heavy nuclei (e.g.
Fe) to protons. A careful study of the highest energy event (the one at 320 EeV,
also detected by the Fly's Eye) seems to exclude only a pure electromagnetic
shower, the profile of the shower being compatible with that generated by a
light or heavy hadron. A neutrino induced hadronic shower remains possible but
this brings up the interaction probability of a neutrino with the higher
atmospheric layers.

A first look at the reconstructed directions of the EHECRs seems to show no
clear pattern indicating the existence neither of pointlike sources nor of large
scale structures where such sources could be grouped. One should keep in mind
that at such energies, and from what we know of the galactic and extragalactic
magnetic fields, the EHECRs should point directly to their sources if these are
reasonably close. The Larmor radius of a charged particle at 320 EeV is larger
than the size of the galaxy if its charge is less than 8. If we take the
currently accepted upper limit ($10^{-9}$ G) for the extragalactic magnetic
fields, a proton of the same energy should have a Larmor radius of 300 Mpc or
more, when the propagation arguments (see below) would exclude its source being
at a distance larger than a few tens of Mpc. 

A recent study \cite{Stanev} using the totality of the data available above 20
EeV shows that there is a quite convincing correlation between the arrival
directions of the subsample of events with energies above 40 EeV and the
supergalactic plane, a structure related to the local group and roughly
perpendicular to the galactic plane. The authors note that this same plane
is also the direction where there is an especially high density of
radiogalaxies. It is premature to consider such an analysis 
as bringing any proof as for the nature of the accelerating mechanisms. Much
larger statistics are needed for such statistical methods to be definitively
conclusive.

Among the recent data published by the AGASA collaboration, there are three
pairs of events whose directions are superimposed. If one \underline{assumes} 
that these pairs come from common pointlike sources\footnote{There is a 2\%
probability that the directions of the three pairs overlap just by chance.}, 
interesting conclusions can be reached without appealing to any particular 
model \cite{Cronin}. For example, just by using
the known values and pattern of the galactic magnetic fields, it can be shown
that the detected cosmic rays cannot be anything except protons (if they are
charged, of course). Then, setting the usual upper limits to the distance of the
source, one can show that the limit of the average extragalactic magnetic field 
(with a hypothesis of 1 Mpc cell size) is about $5\times10^{-10}$ G, a value
very close to that obtained with the help of sophisticated analysis techniques
such as Faraday rotation of polarized light coming from distant galaxies. 
Although one cannot say more than what is stated above about the
common origin of overlapping events, this shows how rich the analysis of such
events can be, provided there is a reasonable amount of them available.

\section{Origin and transportation of the EHECRs}
We made previously the strong statement that the observed EHECRs should not
exist. This needs no doubt some explication and proof. What we mean by this
statement is the following. The observation of such a cosmic ray's interaction
by a ground-based detector means that previously it went through a series of
processes and trials which, all put together, seem almost impossible to satisfy
the observed facts.
One should  use the adverb ``almost" since we know that for a very small
quantity of cases it did indeed happen. First the particle has to be accelerated
at energies much larger than that measured when detected. For this we shall 
consider mainly what we call ``conventional accelerating mechanisms" known in 
astrophysics. Then it has to escape from its production  site~: there one has to
take into account ``beam-dump" effects which exist in the vicinity of most of
the candidate sources and where a substantial part of the particle's energy is
tapped off. Provided there is enough energy left, the particle has then to
travel across the intergalactic medium where inelastic reactions with the cosmic
microwave background (CMB) occur and degrade more of the particle's energy. And
finally the cosmic ray interacts with the atmosphere creating an extensive air
shower (EAS) whose physical characteristics have to be in conformity with what
is observed in the few cases mentioned above. Any proposed scenario attempting
to give an explanation on the origins of the EHECRs has to be checked against
not only one (e.g. the first) of these steps but against the complete set of
them~: only then it can be taken seriously.

Whenever we shall need to illustrate our case by an explicit observation, we'll
use the example of the champion of all observed cosmic rays, which is the one
of 320 EeV detected in 1992 by the Fly's Eye and whose study and interpretation
generated an abundant litterature (see e.g. one of the most detailed of those
articles \cite{Elbert}).

\subsection{The conventional accelerating mechanisms}
By {\it conventional} we mean accelerating mechanisms based on the
electromagnetic fields existing inside or in the vicinity of known astrophysical
objects. Whatever the acceleration mechanism, there is a necessary constraint
which has to be verified by the ``site" where this mechanism operates and which
relates its size to the magnetic field existing there. This condition was
pointed out years ago by M.Hillas \cite{Hillas} and can be understood in the
following way. A particle which is progressively accelerated to the highest
energies has to be contained inside an ``acelerating site" during the whole
process. This is only possible if its Larmor radius is smaller than the size of
the site characterized by a radius $R$. This imposes a simple condition on the
magnetic rigidity, which one can express as an easy-to-remember
{\it approximate} formula~:

$$BR>\frac{E}{Z}$$
where $Z$ is the charge number of the particle, $B$ is in $\mu$G, $R$ in kpc 
and $E$ in EeV. Since this is a necessary condition, one can limit the search
for candidate sources to those where the magnetic rigidity checks this relation.
The popular and well known ``Hillas plot" consists in positioning the known
astrophysical objects in a figure where, in logarithmic scales, $R$ is plotted 
on the abscissa, $B$ in ordinates and diagonal straight lines show the wanted
values of the energy as a function of $Z$. If one looks at sites where a proton
can reach the energy of 100 EeV, one can see that very few possibilities remain
above the corresponding line~: powerful radio-galaxies in the large $R$-low $B$ 
region, neutron stars in the large $B$-low $R$ region and active galactic 
nuclei (AGN) in between. Although in all three cases the acceleration 
mechanism is different, these are the most powerful candidate accelerators in 
the universe. However, there is no general agreement in the community that
any one of these accelerators is able to reach energies in the ZeV range (Z is
for zetta=$10^{21}$ eV), energies necessary if one wants to understand the
existence of cosmic rays detected at 320 EeV. Most predictions usually limit 
the accelerating power of such systems to a maximum of 10 to 100 EeV. 
However, the most optimistic authors do not exclude the possibility that a
proton may be accelerated at energies of 1 ZeV by diffusive shock acceleration
in quasars and radiogalaxies \cite{Sigl}.

Now, suppose that this is possible. Then at least for two of the sites appears,
among others, the problem of the ``beam-dump" encountered by the particle as 
soon as it is accelerated. For the neutron star, this is precisely the
accelerating (moving) magnetic fields themselves~: the particle is expected to
lose most of its energy by synchrotron radiation when crossing these fields on
its escape trajectory. For the AGN, the dump consists in the very intense
radiation fields surrounding the central parts of those engines. At very high
energies, the energy loss in such regions are the result of processes similar to
those we shall describe in the next paragraph.

\subsection{From the source to the earth}
Suppose again that by some kind of miracle, the preceding obstacles are
overcome, and the particle starts its long journey to earth. It has to go
through a medium filled (with a density of about 400 photons/cm$^3$) with the
CMB, photons with an average energy of about $10^{-3}$ eV. If we blue-shift the
photon to the, say, proton's rest system, we see that we are dealing with
inelastic photoproduction processes (photo-pion production) provided the 
proton has an energy in excess
of about $5\times10^{19}$ eV. The back-of-the-envelope formula giving the energy
of the 3K photon in the rest frame of a proton of 100 EeV is~:
$$E(\mathrm{MeV})\approx\frac{25}{\lambda(\mathrm{cm})}$$
which gives a few hundreds of MeV since the 3K photons have millimetric
wavelengths. Such cross-sections have quite large values (fraction of 
millibarn). The mean free path of a 300 EeV proton is about 6 Mpc 
\cite{Halzen2} and each interaction with the 3K photons degrades the proton's
energy by more than 10\%. This means that the proton very quickly loses its
energy until this goes below the inelastic threshold. The consequence is that
whatever the energy with which the proton started, after at most 100 Mpc or so
its energy is found to be less than 100 EeV. In other terms, if the 300 EeV
cosmic ray is a proton, its source cannot be further than a few tens of Mpc.
This is the origin of the spectral GZK cutoff \cite{Aharonian}.

One could object that there is no reason that the EHECRs have to
be protons, which is right. But one can see that the same kind of conclusions
are reached with any kind of particle, except neutrinos. If one considers the
possibility of these cosmic rays to be photons (which hypothesis should not be
excluded {\it a priori}) the interactions with the background radiations 
(including the radio waves) still limit the range very drastically. The pair
creation threshold is around 300 TeV for the incident photon energy and the
mean-free-path of a 320 EeV photon would be 20 Mpc with an asymptotic limit of
about 30 Mpc \cite{Halzen2}. Another way of seeing the problem is a thumb-rule
giving the survival probability for a photon of 300 EeV after a trajectory of
$L$ Mpc~:

$$P(L)\approx\mathrm{e}^{-L/6.6}$$
which yields a probability of less than $10^{-3}$ for a distance larger than 50
Mpc.

Can the EHECRs be heavy nuclei? We already said that the observed events at
$E>100$ EeV do not favour this hypothesis but cannot exclude it either. However,
here again, the propagation processes put very stringent conditions on the
distance to the source. The photodisintegration reactions with the CMB are
expected to strip on average four nucleons per Mpc from the traveling nucleus.
If a heavy nucleus (iron, oxygen) is produced with an initial energy of, say 800
EeV, after less than 10 Mpc the energy of the surviving fragment goes below 300
EeV and reaches the 100 EeV threshold at around 20 Mpc \cite{Aharonian}. The 
clear conlusion is
the following~: if the highest energy cosmic rays observed were heavy nuclei,
they could only be of galactic origin. In this case, even with large charges
(hence small Larmor radii) some anisotropy in their direction should be easily
detectable with a resonable amount of statistics.

The overall conclusion from the arguments above is that the potential source for
the highest energy cosmic rays observed cannot be situated at distances much
larger than a few tens of Mpc (i.e. probably within the local cluster of 
galaxies), except if these cosmic rays are neutrinos (immune to interactions 
with the CMB). Since at such energies, the reconstructed directions should point
clearly toward the source (except for heavy nuclei), it is easy to look for
astrophysical objects in the categories mentioned in the previous paragraph
contained in a sphere of, say, 50 Mpc around us. This exercice was made for the
directions of the highest energy cosmic rays (especially for the 320 EeV Fly's
Eye event) and no candidate source was found.

Can the EHECRs be neutrinos? This is a very interesting hypothesis since it
kills all arguments limiting the distance of the source. Several authors studied
this possibility \cite{Sigl,Elbert,Halzen2}. The answer is ``yes" but with many
question marks. First, the neutrino  induced shower has to reproduce the
features of the observed EAS. For this, the Fly's Eye detector can be powerfully
discriminating since it observes the longitudinal development of the shower and
can trace back the shower's origin. These have very caracteristic features
depending on the initial particle. If the incident particles were neutrinos, not
only would we need very large fluxes to explain the few observed events but also
we would expect those to interact with a flat distribution in atmospheric 
depth.
However one can answer this objection by another~: ``Who knows how a 100 EeV
neutrino actually interacts with nuclei?" If one takes seriously such an
argument, it would mean that the EHECRs open a new window in particle physics.
Unfortunately in this case, we can see no obvious method for distinguishing
neutrino showers from their hadronic or electromagnetic partners. If one stays
within the limits of the standard model, there is an easy way for a ground
detector to discriminate neutrino showers from the rest. We'll say a few words
about the ability of the Auger Observatory to detect EHE neutrinos in the last
paragraph. 

Supposing we ignore the distance constraint for the 320 EeV event (maybe
invoking the neutrino hypothesis). One has still to answer some questions of 
astrophysical origin. As an example, it has been noted \cite{Elbert} that there 
is a very powerful Seyfert galaxy (MCG 8-11-11) in the 
approximate direction of this event but at a distance of 900 Mpc. It is however
impossible to explain why such a source capable of producing a 320 EeV particle 
would give no clear signal (easily visible by the Fly's Eye detector) in the 10
EeV region where the attenuation with the distance even for protons is much
weaker.

Thus the overall comment on what precedes is what was announced in the
beginning~: when all the necessary processes and observational facts have been
taken into account, no globally acceptable explanation is found for the
existence and detection of the EHECRs.

\subsection{The exotic sources}
Before beginning the description of the dedicated experimental project, it is
worthwhile to say a few words about some non-conventional ideas on the possible
sources of EHECRs.

In their search for models explaining large scale structure formation in the
universe, cosmologists have produced a class of appealing models where appear
objects called ``topological defects" \cite{Vilenkin}. The cosmic strings are
considered the most likely of the topological defects to be observed some day.
The production of these defects are initially linked to phase transitions in the
early universe but present day mechanisms producing such objects seem not to be
excluded. A cosmic string is a region of space where huge energy densities
($10^{22}$ g/cm) are present. Such regions could trap grand-unification
particles of mass up to $10^{24}$ eV. In some cases (two crossing strings,
collapse of a string on itself), the GU particles would be liberated. Since they
are unstable, they would immediately disintegrate and emit large numbers of
ordinary particles whose energies could easily exceed those of the EHECRs. Some
authors even envisage that the cosmic string itself could interact with the
atmosphere and initiate the EAS.

This very speculative hypothesis is also very attractive because it is
practically the only one which would survive if it is confirmed that the sources
of the EHECRs are indeed very close and diffuse.

Many authors \cite{Sigl,Gill,Yoshida} checked the cosmic string hypothesis 
against the experimental data. Although some hypotheses in this sector have to 
be abandoned, such as {\it gauge strings} as being responsible of the observed
EHECRs \cite{Gill}, many models remain viable and testable (e.g. by high rates 
of photon component in the EHECRs) and may well open a new window in cosmology 
and particle physics.  

\section{The Auger Observatory}
The present day situation as regards the highest energy part
of the cosmic ray spectrum, the many unknowns as to the origin of these cosmic
rays, the posibility that those events may open new windows in cosmology,
astrophysics or particle physics, all these are strong arguments in favour of a
dedicated experiment. What can one ask of such a project? The first and most
important requirement is that it accumulates statistics. It is hopeless to
expect any element of answer to this astrophysical puzzle if the detection rate
of the relevant events goes on at the present level, which means several more
decades to have a few tens of events above the GZK cutoff. Then, the projected
detector must be able to measure the energy of the incoming cosmic rays with a
better resolution (say, 10\%) than what is available now. It must have a good
resolution in measuring the direction of the cosmic ray since, as was said 
above, we'll be doing astronomy with these particles. And finally, the identity 
of the incident particles (or the chemical composition if those are a mixture of
several species) must be detected on a statistical, if not individual, basis.

The Auger observatory\footnote{Pierre Auger, a french physicist who played an
important role in the founding of CERN, was the first to have the intuition that
cosmic ray events observed coincidentally in distant detectors could be
extensive air showers of very high energy.} is precisely designed to fulfill the
whole set of those requirements. We'll give a brief description of this detector
and of its expected performance, and advise the interested reader to consult a
detailed technical report written by a working group of many people during the
first six months of 1995 \cite{Auger}.

\subsection{Description of the detector}
The Auger observatory is a ``hybrid" ground detector which will be installed on
two sites, respectively in the southern and northern hemisphere. The aim of
having two components to the detector is to be able to see the whole sky. The
southern hemisphere detector is especially interesting since very few detectors
took data in the past in this part of the world from where the direction of the
center of our galaxy is visible. 

The detector will be designed to be fully efficient for showers with energies of
10 EeV and above. This will make the link with the part of the energy spectrum
well explored by presently operating detectors, especially AGASA and Fly's Eye.

Each site will be equipped with an array of ``stations" covering an area of
about 3000 km$^2$. Thus, the total surface of 6000 km$^2$ is expected to yield a
rate of several thousands of events detected per year above 10 EeV, and between 
50 and 100 above 100 EeV. Each giant array will be completed by an optical 
device detecting the fluorescence light emitted by the nitrogen molecules of the
atmosphere excited by the charged particles produced in the EAS. The northern
site should host a detector derived from the japanese Telescope Array
project \cite{Teshima}. The southern site will be equipped with an improved
version of the Utah University's Fly's Eye \cite{Auger,Sokolsky}. The reason for
the detector to be ``hybrid" is that, for about 10\% of the events jointly
detected by the array and the optical device, not only the quality of the data
will be much improved but also each detector will be used to cross-calibrate the
other since they measure independently and with different methods all the
shower's parameters.

The individual stations will be water-Cerenkov tanks where the Cerenkov
radiation emitted by the charged particles penetrating in the detector will be
read out by three phototubes. The output signal will be digitized by rapid
cycling flash ADCs, whose aim is to separate the electromagnetic signal (low
energy photons and electrons) from the muons crossing the tank. The relative
synchronization of the stations (with a few ns precision) will be done using
signals emitted by GPS (Global Positioning System) satellites. The communication
between stations and the data transfer to a central computer will be done by
using radio signals by methods similar to cellular telephone techniques. The
stations will be powered with solar panels and batteries.

The constraints on the site come from the above specifications. The lattitude
should be around $40^{\circ}$ North and South. The array needs
a 3000 km$^2$ flat region (line-of-sight visibility between neighbouring
stations for radio reception) with moderate
temperatures (water tanks) and cloudless skies (solar panels). The optical
detector needs a transparent atmosphere and no source of light pollution nearby.

Among half a dozen sites visited by a survey team in the southern hemisphere,
the region of las Le\~nas (north of Patagonia in Argentina) was chosen by the 
collaboration during a full meeting in November 1995. The northern site will be
chosen in September 1996.

\subsection{The direction of the primary}
In the optical detectors, pixels seeing just a small portion of space record the
time of arrival of light pulses in this direction. The array registers the time
of arrival of the particles in the water tanks. This timing information enables
both detectors (with different methods) to reconstruct the direction of the
shower axis, hence the direction of the incident particle. For showers close to
the vertical, the resolution for the array alone is of the order of $2^{\circ}$
or better above 10 EeV. For hybrid detection of showers, this resolution
can reach $0.2^{\circ}$. 

\subsection{The energy measurement}
On average, a charged particles deposits, for every depth equivalent of 1
g/cm$^2$, an energy of 2.2 MeV. The integral of this energy deposit converted
into fluorescent light is seen by the optical detector as the longitudinal
development of the shower. A particularly important parameter measured by this
detector is the position of the shower maximum $X_{\mathrm{max}}$. The total
detected light and $X_{\mathrm{max}}$ are the two parameters through which the
total energy of the shower can be estimated. In the 10-100 EeV range, the Fly's
Eye energy resolution is of the order of 30\% including statistical fluctuations
and systematic uncertainties.

The ground array usually measures the energy of the shower from the lateral
density distribution of particles. There is a ``magic" parameter coming from
lower energy methodology, the density at 600 m from the shower core, which was
shown to be the best estimator, through a rough proportionality law, of the
incident particle energy. Actually, this distance is simply the one where
fluctuations due to the shower tail and those coming from the lateral
distribution partly compensate and the total fluctuation becomes minimum. The
density at this distance is quite weakly dependent on the nature of the initial
particle. At
higher energies relevant to Auger, this distance seems to be slightly larger.
The estimation of the total energy in Auger should be improved by the fact that
the detector will be able to measure separately the muon and electromagnetic
densities. For hybrid measurements, we expect the energy resolution to be better
than 10\% at 100 EeV.

\subsection{Identification of the primary}
The main feature which differentiates showers coming from heavy or light nuclei
is the depth at which occurs the shower maximum $X_{\mathrm{max}}$. Shower
development studies showed that a shower from a heavy nucleus behaves more or 
less as a superposition of showers coming from the individual nucleons of the
nucleus sharing the total energy between them. This explains why the shower
maximum is at a higher altitude for heavy nuclei and also why those have a
larger muon component at the ground level (the pions reach faster the low energy
limit below which they decay before interacting). Heavier nuclei produce also
faster risetimes (the rate at which the particles reach the ground at a given
distance from the shower core). Those are the three parameters one can use to
identify the primaries~: the time structure of the ground particles, muon to
electromagnetic densities as a function of the distance to the core, the
position of the shower maximum. The first two can be measured by the ground
array, the last by the optical device, hence providing two independent methods
to identify the primary. The simulations show that if the muon to
photon/electron separation in the Cerenkov detector is good enough, the primary
composition should be measured on a statistical basis if not for individual
showers.

\subsection{The detection of neutrinos}
Since this is the main topic of this conference, let us show that, at least for
the highest energies, Auger should perform as a competitive neutrino detector.
The neutrino astronomy above 10 EeV is directly related to the GZK spectral
cutoff~: this is the only remaining hypothesis if one excludes the possibility
that the EHECR sources are actually situated within a few tens of Mpc.

The Fly's Eye group \cite{Yoshida} have studied the possibility for this detector
(the HiRes version) to observe neutrinos from topological defects. The results
depend strongly of course on the incident fluxes, hence of models, but a few
events of this type are expected for a long term experiment. The aperture of 
the HiRes detector for neutrinos with energies of the order of 100 EeV is 
expected to be between $10^4$ and $10^5$ m$^2$.sr. Of course, the detection rate
suffers from the irreducible 10\% duty cycle of any optical detector whose
operation is limited to dark moonless nights and clear skies.

Cronin, Parente and Zas have recently studied the possibility for the ground
array of detecting horizontal showers. Their partial results were presented at
this workshop by E.Zas. Let us just remind that the water Cerenkov tanks are
sensitive to charged particles, whatever their direction. Thus for horizontal
air showers (HAS), provided their axis is at a reasonable altitude (say, less 
than 3 km) above the ground level, the Auger array behaves as a calorimeter 
with uniform density of matter and longitudinal sampling. Showers with 
directions at very large angles, if observed, can hardly be anything except 
neutrino induced (the horizontal atmosphere thickness is roughly 36 times the 
vertical one). The component due to hard muon
bremsstrahlung can easily be eliminated with an appropriate trigger or off-line
analysis. If the neutrino interaction (which should  be uniformly
distributed) happens to be close enough to the array so that the shower maximum
falls within the instrumented area, then the array becomes a very powerful
detector since it samples the totality of the shower components, provided the
neutrino rates are appropriate. Those were studied by the above mentioned
authors and the possibility that the Auger array may detect a few tens of
neutrino induced HAS above 10 EeV per year seems realistic. In any case,
independent of the incident fluxes, one figure to keep in mind is that the Auger
array has an acceptance of 10 km$^3$.sr water equivalent for HAS of 10 EeV, this
acceptance being a slowly increasing function of the energy. 

\section{Conclusion}
The projected Auger observatory is presently in the R\&D phase where all the
technical issues developed in the design report \cite{Auger} are being optimized
and tested. About 150 physicists and engineers from 15 countries are working on
this project just now. Its scientific aims are very clearly established~: this
is an observatory (very much as an astronomical one) optimized for the detection
of cosmic rays with energies above $10^{19}$ eV. Such events, if detected in
large quantities, may enable us to answer one of the very rare (hence very
exciting) open questions in the field of astrophysics and/or cosmology~: what
are the sources of the most energetic particles observed in the universe, i.e.
the most powerful machines existing in Nature? 

Skimming through the steady flow of recent articles related to high energy
astrophysics, one can easily see that this physics issue may have many
implications, i.e. it may be related to a variety of fields where the
(particle) astrophysics community is quite active~: AGNs and pulsars as powerful
accelerating engines, cosmic strings or other topological defects, gamma ray
bursts \cite{Waxman}, neutrino astronomy etc... 

Finally, one should be aware that this project is also open to people not
primarily interested in its specific physics motivation, but who could consider
it as an available experimental ``facility". The Auger observatory provides two
fully equipped areas of 3000 km$^2$ each with possibility of very precise
synchronisation over the whole surface, telecommunication equipments to
communicate and exchange data, a central station equipped with a computer center
and connections to high-flow international networks and real-time access to 
data from any part of the world. The capacity of such a detector to work with
other types of equipments is limited only by the imagination of the potential
users.

\end{document}